\documentclass[twocolumn,amsmath,amssymb,pra]{revtex4}
\usepackage{epsfig}
\usepackage{color}
\usepackage{graphicx}
\usepackage{bm}
\usepackage{epstopdf}

\begin{document}

\title{Repulsively interacting fermions in a two-dimensional deformed trap with spin-orbit coupling}

\author{O. V. Marchukov, D. V. Fedorov, A. S. Jensen, A. G. Volosniev, N. T. Zinner}
\affiliation{Department of Physics and Astronomy, Aarhus University, DK-8000 Aarhus C, Denmark}

\date{\today}

\begin{abstract}
We investigate a two-dimensional system of with two values of the
internal (spin) degree of freedom. It is
confined by a deformed harmonic trap and subject to a Zeeman field,
Rashba or Dresselhaus one-body spin-orbit couplings and two-body
short range repulsion. We obtain self-consistent mean-field $N$-body solutions
as functions of the interaction parameters.  Single-particle spectra
and total energies are computed and compared to the results without
interaction. We perform a statistical analysis for the distributions of nearest
neighbor energy level spacings and show that quantum signatures
of chaos are seen in certain parameters regimes. Furthermore, the effects of two-body 
repulsion
on the nearest neighbor distributions are investigated.
This repulsion can either promote or destroy
the signatures of potential chaotic behavior depending on relative strengths of parameters. Our 
findings support the suggestion that cold atoms may be used to 
study quantum chaos both in the presence and absence of interactions.
\end{abstract}

\maketitle

\section{Introduction}
\label{sec:intro}

With the development of experiments with ultracold atoms, molecules
and ions a lot of models from various areas of physics were given a chance
to be tested in so to speak ``clinical'' conditions~\cite{ketterle2008,bloch2008,esslinger2010,cirac2012}. 
The extreme purity and control of the system parameters allows one to focus only on the effects
in question and not worry too much about defects, impurities, etc. which
always contribute in condensed-matter systems like, for instance, solids.

Not surprisingly these systems attract enormous interest both
from theoretical and experimental groups all around the world. One particularly
interesting area of research is the effect of spin-orbit coupling in
atomic gases. Created with the help of sophisticated optical setups such systems
provide an insight on the effects of the interaction between the spin of a particle and 
its motion. Unfortunately, such experiments are extremely complicated and arbitrary
spin-orbit coupling in ultracold atomic gases has not yet been realized. However, 
several specific cases were achieved in state-of-the-art experiment both for
bosons~\cite{lin2009a,lin2009b,lin2011,aidelsburger2011,zhang2012} and fermions~\cite{wang2012,cheuk2012}.

In experimental setups the atoms are usually trapped by an external
magnetic and/or optical trap, which often can be approximated with a
harmonic potential. The interplay between the trap and the
contribution from the spin-orbit coupling leads to interesting effects
in the energy spectra. Examples of mathematically similar problems are
the quantum Rabi model~\cite{chen2013} and the so-called $E\otimes
\epsilon$ Jahn-Teller model~\cite{larson2009}.  For small values of
the spin-orbit coupling the problem can be treated perturbatively.
However, only even powers of the spin-orbit coupling contributes, that
is first order vanishes and the second order correction is the lowest
non-zero contributing term~\cite{blume2014a}.  Numerical treatment of
the problem shows a peculiar dependence of the eigenlevels on the
value of the spin-orbit coupling strength~\cite{marchukov2013}.  This
dependence corresponds to the so-called Fock-Darwin
spectrum~\cite{avetisyan2012, reimann2002}.

A non-interacting system in two spatial
dimensions confined by a deformed harmonic trap with spin-orbit
coupling of Rashba \cite{rashba1960} and Dresselhaus \cite{dresselhaus1955}
type and a Zeeman field has stability properties that depend
sensitively on these external parameters \cite{marchukov2013}. 
Furthermore, by analyzing the statistics of energy level spacings
and their distributions, one may infer that these systems may
give rise to chaotic dynamical motion \cite{marchukov2014a}.
For a given number of fermions the level structure around
the Fermi energy is crucial for several properties of the system.  In
particular, the density of states is directly related to
stability. It can therefore be expected that including interactions
would exhibit quickly changing properties as function of a
repulsive two-body interaction which in turn would produce varying
single-particle density at the Fermi energy.  To include such an
interaction a commonly employed procedure is to calculate the average
effects self-consistently in the mean-field approximation. Using 
this approach we can then address how the dynamics may be
influenced by interactions by statistical analysis of the nearest 
neighbor energy level spacing distributions in analogy to the 
non-interacting case \cite{marchukov2014a}. The Hamiltonian 
we study here has resemblance to molecular physics studies of 
the Jahn-Teller model \cite{larson2009} and the question of irregular
and chaotic dynamics has been discussed previously 
in that context \cite{markiewicz2001,yamasaki2003,majernikova2006a,majernikova2006b}.

The purpose of the present paper is to report on the effects of a
short-range two-body interaction in addition to the externally
controlled single-particle fields for an $N$-body system
of fermions.  The external parameters active in two spatial
dimensions are two frequencies in a deformed harmonic trap, magnetic
field strength for Zeeman splitting, Rashba and Dresselhaus spin-orbit
strengths.  The repulsive two-body interaction is a delta-function with a
variable strength. We note that the interplay between the short-range interaction
and the spin-orbit coupling in few-body systems was recently discussed in
Refs.~\cite{blume2014a,blume2014b,cui-prx2014}.

In Section~\ref{sec:formalism} we present the
theoretical framework with notation and corresponding definitions.
The calculated single-particle spectra and the total energies are
discussed in Section~\ref{sec:results} and compared to the non-interacting case.  
As indicative for the dynamical
behavior we also analyze the statistical properties of
the spectra as function of the repulsive two-body interaction in Section~\ref{sec:results}.  
Finally, Section~\ref{sec:conclusion} contains a brief summary and our conclusions.

\section{Formalism}
\label{sec:formalism}
In this section, we first
specify the Hamiltonian with one- and two-body potentials, then derive
the mean-field equations, and discuss the (possibly different)
symmetries of both the fully correct solutions and the mean-field
approximation.

\subsection{Hamiltonian}
\label{subs:Hamiltonian}

We consider a system of $N$ identical spin-$\frac{1}{2}$ fermions confined to two spatial
dimensions (2D) by a one-body harmonic potential.  In the experiments 
with the ultracold alkali and alkaline atoms the two available values
of the internal degree of freedom correspond to different hyperfine
states of an atom ~\cite{bloch2008}.  This is formally described as a
particle of spin-$1/2$ with two possible projections. 
The Hamiltonian includes the spin-dependent one-body spin-orbit coupling
and the two-body interaction terms. It is given by
\begin{eqnarray}
\label{Hamiltonian}
 \hat H & = & \sum_{i=1}^{N} \hat H_{0}(i) 
 +   \sum_{i < j} \hat V_{ij}  \; , \\
 \hat H_{0}(i) & = &
\left ( \frac{\mathbf {p_i^2}}{2m} + \frac{1}{2} m (\omega_x^2 x_i^2 + \omega_y^2 y_i^2) \right ) \otimes \hat I 
 + h \hat \sigma_{iz} \nonumber  \\ & + &    
  (\alpha_R + \alpha_D){\hat \sigma_{ix}} {p_{iy}} - 
(\alpha_R - \alpha_D) {\hat \sigma_{iy}} {p_{ix}} \; ,
\label{sphamil} \\ 
 \hat V_{ij} & = & \hat P_0 V(\mathbf r_i - \mathbf r_j)\; ,
\label{inter}
\end{eqnarray}
where $m$ is the mass of one particle, $\mathbf r_i = (x_i, y_i)$ and
$\mathbf p_i = (p_{ix}, p_{iy})$ are 2D coordinates and momenta of the
$i$'th particle, $\omega_{x}$ and $\omega_{y}$ are the frequencies of
the possibly deformed harmonic trap, $\hat I$ is the $2\otimes 2$ unit
matrix. From now on a hat denotes that the value is a $2\otimes 2$ matrix.

The spin-dependence is given in terms of the Pauli matrices, $\hat
\sigma_{ix}$, $\hat \sigma_{iy}$ and $\hat \sigma_{iz}$. $h$ is the external
Zeeman field, $\alpha_R$ and $\alpha_D$ are the strengths of the
Rashba ~\cite{rashba1960} and Dresselhaus ~\cite{dresselhaus1955} spin-orbit couplings (SOC), respectively.
The strengths, $\alpha_R$ and $\alpha_D$, have dimensions of velocity
which for an oscillator is measured by $v_{osc} = \sqrt{\frac{2\hbar\omega_y}{m}}$.
Throughout the paper, $\omega_y$ will be our reference frequency, i.e. we will 
measure $\omega_x$ in units of $\omega_y$.
We shall present the results using this unit. The two-body interaction is repulsive,
with $V(\mathbf{r}_i - \mathbf{r}_j)$ as a spatial part of the interaction 
and the operator, $\hat P_0 = (\hat I - \boldsymbol{\hat \sigma}_i \cdot \boldsymbol{\hat \sigma}_j)/4$,
which projects on singlet states. As we discuss below, the interaction term will be short-range
and we model it by a delta-function (pseudo)-potential. In this case the Pauli principle excludes
effects of interactions in the triplet channel as we are considering fermions.

\subsection{The Hartree-Fock equations}

The mean-field, or Hartree-Fock, approximation for a system of
identical fermions consists in finding the lowest energy for a fully
antisymmetrized product, $\Psi_{HF}$, of single-particle wave functions.  This product
is expressed as a Slater determinant, that is
\begin{equation}
\label{psi_hf}
\Psi_{HF} = \frac{1}{\sqrt {N!}}
\begin{vmatrix}
\psi_1(\mathbf r_1) & \psi_2(\mathbf r_1) & \cdots & \psi_N(\mathbf r_1) \\
\psi_1(\mathbf r_2) & \psi_2(\mathbf r_2) & \cdots & \psi_N(\mathbf r_2) \\
  \vdots  & \vdots  & \ddots & \vdots  \\
\psi_1(\mathbf r_N) & \psi_2(\mathbf r_N) & \cdots & \psi_N(\mathbf r_N) 
\end{vmatrix},
\end{equation}
where $\psi_i$, ($i = 1, \ldots, N$) are single-particle wave
functions which in the present case describe two-component (spin-up and
spin-down) states.  As a vector with two spinor components we
write it as $\psi_i(\mathbf r) = (\psi_{i\uparrow}(\mathbf
r),\psi_{i\downarrow}(\mathbf r))^{\dagger}$.  The index $i$ labels
the quantum numbers necessary to specify the corresponding state. To
have a non-trivial Slater determinant there must be $N$ linear independent
single-particle wave functions. The states are normalized with the
following condition: $\int \mathrm{d}\mathbf{r} \psi^{\dagger}_i \psi_j =
\delta_{ij}$.  In order to find the approximation to the
total many-body energy we minimize the energy functional\begin{equation}
E[\Psi_{HF}] = \frac{\langle \Psi_{HF} \mid \hat H \mid \Psi_{HF}
  \rangle}{\langle \Psi_{HF} \mid \Psi_{HF} \rangle} \; ,
\end{equation}
where $\Psi_{HF}$ and $\hat H$ are given in eqs.~\eqref{psi_hf} and
~\eqref{Hamiltonian}, respectively.  The minimization with respect to
independent real and imaginary parts of the single-particle wave
functions gives the self-consistent set of
Hartree-Fock equations~\cite{jensen1987}
\begin{eqnarray}
\label{hf_eqs} &&
\varepsilon_i \psi_i(\mathbf r) =  \hat H_{0}(i) \psi_i(\mathbf r) 
\\  \nonumber 
 &+&\sum_{j=1}^N \int \mathrm{d} \mathbf{r'} \psi_j^{\dagger}(\mathbf r') \hat V_{ij} 
 \left [ \psi_i(\mathbf r) \psi_j(\mathbf r') - \psi_j(\mathbf r) \psi_i(\mathbf r') \right ] \; , 
\end{eqnarray}
where $\hat H_{0}$ is the non-interacting single-particle part of the
Hamiltonian in eq.~\eqref{sphamil}, and $\varepsilon_i$ become
single-particle energies, although they are formally introduced as
Lagrange multipliers to maintain normalization.

To continue we need to specify the interaction term.  In experiments
the interparticle interaction of neutral non-polar
cold atoms originates from the van der Waals interaction and is of very short
range. Therefore a convenient parametrization often used is to  
described it by the Dirac delta-potential, that is 
$V(\mathbf r - \mathbf r') = g \delta(\mathbf r - \mathbf r')$.
Here the strength, $g = g_s \hbar^2/m$,
where $g_s$ is dimensionless such that $\hat V$ maintains dimension of
energy. However, applying such a (pseudo)-potential must be done
with considerable care to avoid inconsistencies \cite{olshanii2002,valiente2012a}.
This is particularly important in low-dimensional setups \cite{petrov2000,petrov2001,valiente2012b}.
In the present paper we will consider the weakly interacting limit corresponding to 
the case where the three-dimensional scattering length, $a_{3D}$, is much smaller than 
the transverse confinement length, $l_z$ that is applied to reduce our system to 2D, i.e.
$l_z\gg |a_{3D}|$. In this case $g$ becomes proportional to $a_{3D}$ \cite{petrov2000}, or
$g_s\propto a_{3D}/l_z$ (up to factors of order one), and thus the weakly-interacting 
regime is $g_s\ll 1$. However, in order to investigate potential effects of going to stronger
interactions we will push to larger values $g_s\leq 1$. While this is pushing the 
regime of validity of the pseudo-potential, we show that our mean-field results
can be accurately reproduced by perturbation theory which indicates that the 
strongly interacting regime is beyond $g_s=1$. We therefore do not expect
that including a better approximation for the pseudo-potential will have 
any qualitative effects on our results.

We can insert the interaction~\eqref{inter} into the equations~\eqref{hf_eqs},
with $V(\mathbf r - \mathbf r') = g \delta(\mathbf r - \mathbf r')$ and
the projection operator acting on the spinors $\psi_i(\mathbf r) \psi_j(\mathbf r')$
in the following way
\begin{eqnarray}
\label{pr-op}
&& \hat P_{0}
\psi_i(\mathbf r) \psi_j(\mathbf r') =  \\  \nonumber
&&  \frac{1}{2} (\psi_{i\uparrow}(\mathbf r) \psi_{j\downarrow}(\mathbf r') -
 \psi_{i\downarrow} (\mathbf r) \psi_{j\uparrow}(\mathbf r')) (\mid\uparrow \rangle_i \mid \downarrow \rangle_j -
  \mid\downarrow \rangle_i \mid \uparrow \rangle_j) \; ,
\end{eqnarray}
where $\mid\uparrow \rangle = \begin{pmatrix} 1 \\ 0  \end{pmatrix}$ and $\mid\downarrow \rangle = \begin{pmatrix} 0 \\ 1  \end{pmatrix}$. The Hartree-Fock equations ~\eqref{hf_eqs} can now be worked out in
detail and written as
\begin{eqnarray}
&\hat H_{0}(i) \begin{pmatrix} \psi_{i\uparrow} \\ \psi_{i\downarrow}  \end{pmatrix} 
+ \frac{g}{2} \Big [(n_{\downarrow} \psi_{i\uparrow} 
- n_{\downarrow \uparrow} \psi_{i\downarrow}) \begin{pmatrix} 1 \\ 0  \end{pmatrix} 
+ \nonumber  \\& + (n_{\uparrow} \psi_{i\downarrow} - n_{\uparrow \downarrow}
\psi_{i\uparrow}) \begin{pmatrix} 0 \\ 1  \end{pmatrix} \Big ] 
= \varepsilon_i \begin{pmatrix} \psi_{i\uparrow} \\ \psi_{i\downarrow} \end{pmatrix},
\end{eqnarray}
where we defined the density matrices
\begin{eqnarray}
\label{densit1}
n_{\uparrow} &=& \sum_{j=1}^{N} |\psi_{j\uparrow}|^2\;, \;
n_{\downarrow} = \sum_{j=1}^{N} |\psi_{j\downarrow}|^2\;, \\ \label{densit2} 
n_{\uparrow \downarrow} &=& \sum_{j=1}^{N} \psi^{\ast}_{j\uparrow} \psi_{j\downarrow} \;, \;
n_{\downarrow \uparrow} = \sum_{j=1}^{N} \psi^{\ast}_{j\downarrow} \psi_{j\uparrow} \;.
\end{eqnarray}
Now we can write down the Hartree-Fock equations in the ordinary matrix form
\begin{equation}
\label{m_hf}
 \bigg( \hat H_{0}(i) - \varepsilon_i \hat I\bigg)
\begin{pmatrix}
\psi_{i\uparrow} \\
\psi_{i\downarrow}
\end{pmatrix}
 =
\begin{pmatrix}
 - \frac{g}{2} n_{\downarrow} &   \frac{g}{2} n_{\downarrow\uparrow}\\
  \frac{g}{2} n_{\uparrow\downarrow} & - \frac{g}{2} n_{\uparrow} 
\end{pmatrix}
\begin{pmatrix}
\psi_{i\uparrow} \\
\psi_{i\downarrow}
\end{pmatrix}.
\end{equation}
This matrix is hermitian, since $n_{\uparrow \downarrow} =
n^{\ast}_{\downarrow \uparrow}$, but not necessarily real.  The
Hartree-Fock equations maintain the usual interpretation as the one-body
Schr{\"o}dinger equation for each particle where the potential, in
addition to the one-body external part, includes an interaction of a particle
with the density of all others. The interaction is then depending
on the wave functions of all other particles as formulated through the
densities of eq.~\eqref{m_hf}. Thus, this total mean-field potential has to be 
self-consistently determined through the state of all the interacting particles.

For relatively small interaction the main contribution to the energy
of the system comes from the external potentials. This part is by
definition independent of the two-body interaction and describes the
non-interacting system ($g=0$) with the corresponding one-body
Schr{\"o}dinger equation for a spin-orbit coupled particle in a
deformed harmonic trap and a Zeeman field.  In this limit, the set of
eigenvalues, $\varepsilon_i$, are directly seen to be the
single-particle energies. This single-particle interpretation is also
applied after inclusion of the self-consistently defined contributions
from the interaction term \cite{jensen1987}.  The total energy of the
system can be written in two ways, that is
\begin{eqnarray}
\label{totEnergy}
E[\Psi_{HF}] &=& \sum_{i=1}^{N}\langle \psi_{i}|\hat H_{0}| \psi_{i} \rangle +  \frac{g}{2} 
\int \mathrm{d} \mathbf{r} (n_{\downarrow} n_{\uparrow} - |n_{\downarrow \uparrow}|^2) 
 \nonumber \\ \label{energy}
&=& \sum_{i=1}^{N} \varepsilon_i  -  \frac{g}{2} \int \mathrm{d} \mathbf{r} 
(n_{\downarrow} n_{\uparrow}
- |n_{\downarrow \uparrow}|^2)  \;.\;
\end{eqnarray}
In the non-interacting case the result is simply the sum of all
single-particle energies, $\varepsilon_i$, of occupied states.  With
a two-body interaction, this sum of single-particle energies includes
the contributions from each of the particles $i$ interacting with all
other particles. This means double counting for two-body interactions
and half of the corresponding interaction energy must be subtracted to
get the correct total energy.

\subsection{Rotational symmetry and parity}
\label{subs:parity}
Appearance or absence of symmetries and degeneracies are crucial
information for understanding both  stability, dynamic behavior, and
choice of an efficient numerical procedure.  These properties are
related to classical constants of motion and conserved quantum numbers
in quantum mechanics, where the latter in turn are found through operators
commuting with the Hamiltonian.  The possible spatial symmetries are
rotation around one or more axes, and reflection in planes or points. 

The oscillator trap is always invariant under independent rotations by $\pi$ around the
$x$, $y$ and $z$-axes.  The action of $(x,y) \rightarrow (-x,-y)$,
that is the parity operation, $\hat P$, in two dimensions leaves the
oscillator unchanged while changing sign on the spin-orbit terms.
Rotational symmetry only occurs around the $z$-axis when
$\omega_x=\omega_y$.  A general symmetry arises, since the operator
$\hat \Pi \equiv \hat \sigma_{z} \hat P$ commutes with all the terms of the
Hamiltonian in eq.~\eqref{Hamiltonian}, see ~\cite{hu2012a}.  This
holds for all the external parameters and for the zero-range two-body
interaction as well, for it only depends on the vector difference between
two coordinates.

The spin-orbit coupling does not commute either with $\hat \sigma_z$
or with the orbital angular momentum, $L_z$, but
the cases of pure Rashba $\alpha_R (\hat \sigma_x  p_y - \hat \sigma_y p_x)$
and pure Dresselhaus $\alpha_D (\hat \sigma_x  p_y + \hat \sigma_y p_x)$ terms
commute with the operators $L_z + \frac{1}{2} \hat \sigma_z$ and $-L_z + \frac{1}{2} \hat \sigma_z$,
respectively. Indeed, 
\begin{eqnarray}
\label{commut1} 
 &&[ \pm L_{z} + \frac{1}{2}\hat \sigma_{z},\hat \sigma_{x} p_{y}]
= i(\hat \sigma_{y} p_{y} \mp \hat \sigma_{x} p_{x})\;, \\
 && [\pm L_{z} + \frac{1}{2}\hat \sigma_{z}, \hat \sigma_{y} p_{x}]
= i(\pm \hat \sigma_{y} p_{y} - \hat \sigma_{x} p_{x}) \;
\label{commut2}
\end{eqnarray}
and
\begin{eqnarray}
\label{commut3} 
 &&[L_{z} + \frac{1}{2}\hat \sigma_{z}, \alpha_R(\hat \sigma_{x} p_{y} - \hat \sigma_{y} p_{x})]
= 0\;, \\
 && [-L_{z} + \frac{1}{2}\hat \sigma_{z},  \alpha_D(\hat \sigma_{x} p_{y} + \hat \sigma_{y} p_{x})]
= 0 \;.
\label{commut4}
\end{eqnarray}

The ``mixed'' case of both finite $\alpha_R$ and $\alpha_D$ does not have
this symmetry.

The energy spectra of pure Rashba and Dresselhaus spin-orbit couplings
Hamiltonians are the same and so in the remaining part of the paper we
will only consider the pure Rashba case. Then the $z$-projection
of the total angular momentum, $L_z + \frac{1}{2} \sigma_z$, is a good quantum number
for a cylindrically symmetric trap ($\omega_x = \omega_y$).

\subsection{Time-reversal symmetry}
\label{subs:TRS}

Time-reversal symmetry is very important in odd-spin systems
due to the Kramers degeneracy ~\cite{landau1977, sakurai1994} as well as
providing a label on the single-particle solutions. For particles with
spin-$\frac{1}{2}$ the time-reversal operator is $\hat T = i \hat \sigma_y
\hat K$, where $\hat K$ is the complex conjugation operator
~\cite{sakurai1994}.  Then $\hat T$ commutes with the Hamiltonian in
eq.~\eqref{Hamiltonian} provided the Zeeman field is absent, $h=0$, but the
scalar two-body potential may still be present.  This implies that the
solutions for $h=0$ are at least doubly degenerate due to the Kramers theorem.

The Hartree-Fock mean-field Hamiltonian does not necessarily have the
same symmetry as the non-interacting one.
Spontaneous symmetry breaking can occur when it is energetically
favorable and allowed in the numerical iteration procedure.  In matrix
form the commutator of $\hat T$ and $\hat H$ from eq.~\eqref{m_hf} can be written
\begin{align}
\label{tr_hf}
&\left [\hat H_{0},\hat T\right] +
\left[
\begin{pmatrix}
 \frac{g}{2} n_{\downarrow}  &  - \frac{g}{2} n_{\downarrow\uparrow}\\
- \frac{g}{2} n_{\uparrow\downarrow} &  \frac{g}{2} n_{\uparrow} 
\end{pmatrix},
\begin{pmatrix}
0 & 1 \\
-1 & 0
\end{pmatrix}
\hat K \right ]
 \\& \nonumber =
\begin{pmatrix}
g n_{\downarrow\uparrow} & \frac{g}{2} (n_\uparrow - n_\downarrow) + 2h\\
 \frac{g}{2} (n_\uparrow - n_\downarrow) + 2h & -g n^{\ast}_{\downarrow\uparrow} 
\end{pmatrix}. 
\end{align}
We see that for $h=0$ and $g=0$ the commutator vanishes as expected.
However, $h=0$ and finite $g$ in general can preserve the
time-reversal symmetry.  This is seen from eqs.~\eqref{densit1} and
~\eqref{densit2} where $n_{\uparrow} = n_{\downarrow}$ and
$n_{\downarrow\uparrow} = 0$, provided all wave functions can be
chosen to be real. If these restrictions are lifted these density
identities and consequently the time-reversal symmetry can be
violated. Still for the finite value of $h$ the off-diagonal elements
could equal zero and the commutator could vanish. However, it requires
quite peculiar choice of the parameters values and we do not consider
this case in our calculations.

The time-reversal symmetry can always be imposed on the solution, but
a symmetry breaking solution of lower energy may then also exist.
This would be unavoidable for an odd number of particles where the
double degeneracy has to be broken for at least one pair of
single-particle levels.  For an even number of particles it is much
more frequent to find time-reversal symmetry in the lowest energy
Hartree-Fock solution.  An exception could be when accidental
crossings occur of two doubly degenerate levels at the Fermi energy.
It may then be numerically advantageous to split these four levels and occupy two
non-time reversibly symmetric single-particle states.

However, in this paper the self-consistent
two-body interaction term is initially constructed from the set of eigenstates
of the non-interacting system which preserves the time-reversal symmetry.
It means that the eigenstates of the Hartree-Fock Hamiltonian~\eqref{m_hf}
are not allowed to break the time-reversal symmetry as well. Hence, in our
numerical procedure we ensure that the conditions $n_{\uparrow} = n_{\downarrow}$
and $n_{\downarrow\uparrow} = 0$ are maintained. 

Thus, the Hartree-Fock solutions depend only on one independent
density functional, $n_{\uparrow}$, for even $N$ and conserved
time-reversal symmetry.  Then the effect of the two-body interaction
amounts to an additive term proportional to the total density of the
system.  A large effect is then equivalent to a large change of
density, which in turn therefore must differ substantially from the
other contributing one-body parts of the Hamiltonian.  A dominating
interaction contribution then requires a large strength, $g$, compared
to the one-body harmonic oscillator energy.  we shall not consider
such large strengths in this paper.

\section{Numerical results}
\label{sec:results}
In this section we present our numerical results as obtained by
solving the Hartree-Fock equations for different one- and
two-body interaction parameters.  We first indicate the rather
straightforward method employed in the iteration procedure. Then we
offer a perturbative treatment for small two-body strengths.
The resulting Hartree-Fock spectra are compared with the results of
both non-interacting systems and a perturbative treatment of the
pair-potential.  The change in total energy due to the repulsion is
discussed.  Finally, we perform a statistical analysis on the
single-particle spectra in order to classify the dynamical behavior as
either chaotic or regular or perhaps as a complicated mixture.

\subsection{Procedure and approximations}
\label{subs:proc}

The Hartree-Fock equations are like a set of Schr{\"o}dinger equations with density
dependent potentials.  The density, $n_{\uparrow}$, depends on the solution and 
the equations must be solved self-consistently.  Thus, a given density
produces a potential which in turn has single-particle solutions
adding up to the initial density used to produce the potential.  For $N$
identical fermions the solution consists of $N$ orthogonal
single-particle states with corresponding energies.  Therefore it is
advantageous to use a method where the lowest $N$ eigenstates are
simultaneously obtained.

We expand $\psi_{i \uparrow}$ and $\psi_{i \downarrow}$ on 2D harmonic
oscillator eigenfunctions and find the coefficients of the expansion
and the eigenvalues by diagonalization.  The external deformed trap
suggests a correspondingly deformed basis.  However, the spin-orbit
terms are not optimized by the same choice since they couple to
higher-lying cylindrical oscillator shells.  Still, in this paper we
use the deformed Cartesian basis adjusted to the external oscillator
potential.  The basis states are terminated by maximum quantum
numbers, $n_{xmax}$ and $n_{ymax}$ in the $x$ and $y$ such that $\hbar
\omega_x n_{xmax} = \hbar \omega_y n_{ymax}$, where the oscillator
frequencies vary with deformation.  The total number of basis states
in our calculations is larger than $250$, which is more than
sufficient for the relatively small number of occupied single-particle
levels.  Calculations for the statistical analysis procedure also
require a number of unoccupied levels.  For this purpose we used at
least $575$ basis states, depending somewhat on the deformation. The
absolute error is always not larger than $10^{-4} \hbar \omega_y$ for all the single-particle energies
we employed in the calculations.

The basic ingredients are the matrix elements of the Hartree-Fock
potential which consists of one- and two-body terms.  The one-body
pieces, including the kinetic energy operator, are computed by the
straightforward analytic or numerical integration as in
ref.~\cite{marchukov2013}.  Each of the two-body matrix elements is
an integral over two basis functions multiplied by one of the
densities. This can be reduced to a double sum over integrals of
products of four basis functions.  These basic integrals are
calculated numerically with the help of the Gauss-Hermite
quadratures approximation~\cite{abramowitz1972} and stored for use in the subsequent
calculations.  The interaction matrix elements are now obtained by
summing over these basic matrix elements weighted by the expansion
coefficients of the single-particle wave functions.

All matrix elements are combined to give the full Hartree-Fock matrix,
which by diagonalization produces eigenenergies and eigenfunctions.
These eigenfunctions yield a new set of densities which are used to construct a new Hartree-Fock
Hamiltonian where the solutions are either unchanged or inserted in
yet another step of this iterative procedure.  Convergence to
self-consistency is usually achieved after relatively few iterations,
obviously depending on the set of initial wave functions.  Since we
vary at least one continuous parameter, like the spin-orbit coupling
strength, we choose the converged solution as initial
guess for a slightly different strength.  This choice substantially
reduces the number of iterations, and thus speeds up the computations.

The main focus of this paper is on the effect of the two-body repulsion. It is
then interesting to compare the full self-consistent solutions with
the results of perturbative treatments.  Thus, we first compute the
$g=0$ unperturbed solution from $\hat H_{0} (i) \psi^{(0)}_i =
\varepsilon^{(0)}_i \psi^{(0)}_i$. The result has time-reversal
symmetry when $h=0$, and otherwise violates this symmetry.  Then we
write the wave function as $\psi_i = \psi^{(0)}_i + \delta \psi_i$,
where $\delta \psi_i$ is assumed to small compared to $\psi^{(0)}_i$.  The
energy $\varepsilon_i$ is then $\varepsilon_i = \varepsilon^{(0)}_i +
\delta \varepsilon_i$, where $\delta \varepsilon_i$ is a small correction
to the single-particle energy.

The lowest order perturbation theory is given by
\begin{equation}
\label{pertenergy}
\delta \varepsilon_i = \frac{g}{2}
 (\psi^{(0)*}_{i\uparrow},\psi^{(0)*}_{i\downarrow})
\begin{pmatrix}
 n_{\downarrow} &  - n_{\downarrow\uparrow} \\
 - n_{\uparrow\downarrow} &  n_{\uparrow}
\end{pmatrix}
\begin{pmatrix}
\psi^{(0)}_{i\uparrow} \\
\psi^{(0)}_{i\downarrow}
\end{pmatrix}.
\end{equation}
When $h=0$, the densities formed from the Hartree-Fock solutions
fulfill the two identities, $n_\uparrow = n_\downarrow$ and
$n_{\uparrow\downarrow} = 0$, independent of the two-body strength,
$g$.  The perturbation is therefore simply given by
\begin{equation}
\label{perth0}
\delta \varepsilon_i = \frac{g}{2} \int 
( |\psi^{(0)}_{i\uparrow}|^2 + |\psi^{(0)}_{i\downarrow}|^2)
 n_\uparrow  \mathrm{d} \mathbf{r}.
\end{equation}
This approach is equivalent to a perturbative 
treatment of the many-body Schr{\"o}dinger equation.

We can attempt very rough estimates of the energy shift, $\delta
\varepsilon_i$, which would provide a criterion for validity of the
lowest order perturbation approximation.  This requires information
about the $i$'th wave function and the density functions.  Let us assume
that $h=0$ and use eq.~\eqref{perth0}. The density arise from a
summation over all occupied single-particle states.  With harmonic
oscillator solutions we can calculate the root mean square radius,
$R_0$, for a given number of particles, $N$.  Assuming that the
density is constant inside the radius, $R_0$, we have an estimate of
this constant density, $n_0$.  An average value of the perturbation
potential is then obtained to be $g n_0/4$, where we used that the
spin-up density is equal to half of the total density, $n_0$.  
When all the wave functions contributing to the density also are
inside $R_0$, the normalization of the wave functions in
eq.~\eqref{perth0} then provides an estimate of $\delta \varepsilon_i$
equal to $g n_0/4$.

The mean square radius is found for a two-dimensional oscillator
with occupied levels up to $n_f$, that is
\begin{equation}
\label{radius0}
 N R_0^2 = \sum_{i\in occ} <i|r^2|i> = b^2 \sum_{n=0}^{n_f} (n+1)^2
 \approx \frac{1}{3} b^2 n_f^3 \;,
\end{equation}
where $b^2 =\hbar/(m\omega)$, and the degeneracy of each oscillator
energy is $n+1$.  Together with $N= \sum_{n=0}^{n_f} (n+1) \approx
n_f^2/2 = n_0 \pi R_0^2$ we then get $n_f \approx \sqrt{2 N}$ and
\begin{equation}
\label{vpert}
\delta \varepsilon_i \approx g n_0/4 \approx \hbar \omega g_s
3/(4\pi) \sqrt{N/8} \equiv V_{pert}.
\end{equation}
This is now a crude average
estimate of the shift of all levels which then is a state independent
constant.

It is more revealing to compare the size of the average perturbation,
$g n_0/4$, to the size of the controlling harmonic oscillator
potential at an appropriate distance.  If we choose an average
distance of half the root mean square value we get $V_{osc}(r=R_0/2) =
\hbar \omega \sqrt{2N}/12$.  The perturbation is then small in this
unit when $V_{pert}/V_{osc} \approx g_s 9/(4\pi) < 1 $ or when 
$g_s < 1.2$.

When $h \neq 0 $, the double degeneracy of the single-particle
energies is lifted but the perturbation expression in
eq.~\eqref{pertenergy} is still valid.  The $h$-dependence of the
energy correction is then hidden in the unperturbed wave functions.
One overall effect would necessarily again be a shift of all energies
in the spectrum but likely less systematic.

\subsection{Hartree-Fock spectra and $N$-body energies}
\label{subs:single-particle}
The single-particle states contain all the information in the
mean-field calculation and are the most
direct quantities to study. These energies depend on both one- and
two-body parameters, and as well on particle number $N$ for
interacting particles.  The spin-orbit coupling,
$\alpha_R/v_{osc}$, is our choice to exhibit the principal
dependencies, and consequently we select particular values of
deformation, Zeeman strength, two-body strength, and particle number.

\begin{figure}
\centering
\includegraphics[width=\linewidth, height = 8.5cm]{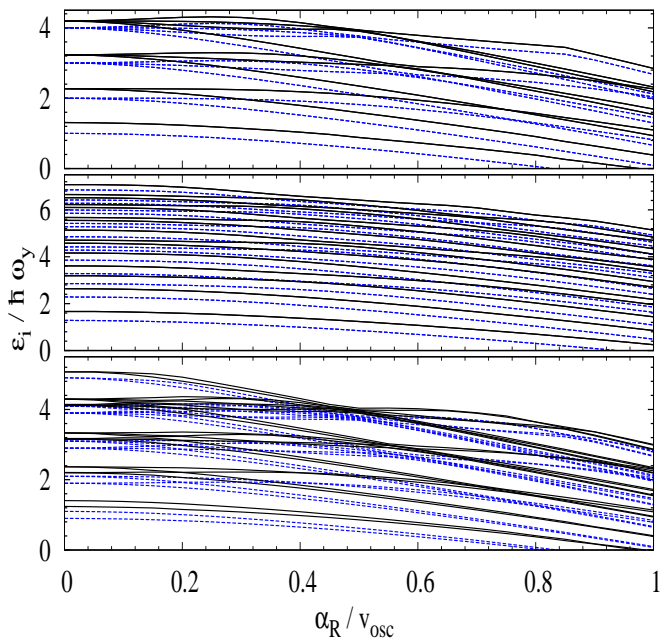}
\caption{The self-consistent single-particle energy levels,
  $\varepsilon_i$, divided by $\hbar \omega_y$ are compared for interacting $g_s=0.5$ (black solid) and
  non-interacting $g_s=0.0$ (blue dashed) particles. 
  The parameters are $N = 20$, $\omega_x = \omega_y$, $h=0$ (upper panel), $N =
  30$, $\omega_x = 1.57 \omega_y$, $h=0$ (middle panel), and $N = 22$,
  $\omega_x = \omega_y$, $h=0.1$ (lower panel).} 
\label{figI}
\end{figure}

\begin{figure}
\centering
\includegraphics[width=\linewidth, height = 8.5cm]{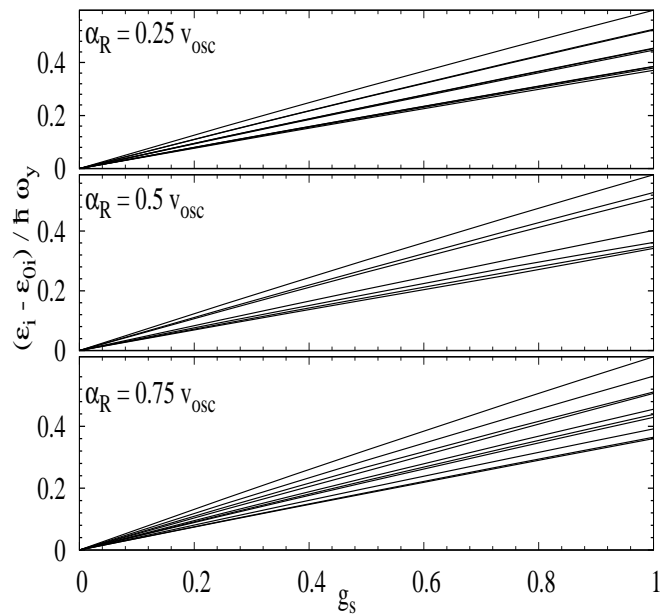}
\caption{The difference between the single-particle energy levels (divided by $\hbar \omega_y$) of interacting and
  non-interacting systems as function of the dimensionless interaction strength
  $g_s$.  The number of particles is $N = 20$, the harmonic trap is cylindrical 
  ($\omega_x = \omega_y$), the Zeeman strength is $h=0$.  The upper, middle and lower panels
  are for $\alpha_R = 0.25, 0.5, 0.75$ in the units of
  $v_{osc} = \sqrt{\frac{2\hbar\omega_y}{m}}$,  respectively. } \centering
\label{figIa}
\end{figure}

In fig.~\ref{figI} we compare sets of $\varepsilon_i$ for
interacting and non-interacting particles, where the latter case was
also discussed in Refs.~\cite{marchukov2013,marchukov2014,marchukov2014a}.  
The upper panel displays the simplest case of a
cylindrical oscillator ($\omega_x = \omega_y$) and no Zeeman field. We choose the particle number $N=20$
since it completely fills the first $4 \times 2$ degenerate shells (oscillator
quantum number is 4) for $\alpha_R=0$. The time-reversal double degeneracy
is present for all levels but the oscillator degeneracy is lifted for
finite $\alpha_R$. The overall behavior is that the repulsion shifts all
energies upwards by an amount roughly independent of spin-orbit
coupling, but with a tendency of decreasing the shift as
$\varepsilon_i$ increase.  The usual avoided crossings appear with a
tendency to be more pronounced for finite $g_s$ in the regions
of high energy levels density as seen for $\alpha_R \simeq 0.5 v_{osc}$.

In the middle panel of fig.~\ref{figI} we start with an ``almost
irrational'' frequency ratio to break as many geometrical degeneracies
as possible by deforming. The particle number is also increased to
$N=30$ to see more occupied levels.  The overall behavior is now at
first glance more complicated due to more levels. However, the shifts
of the energy levels due to the repulsion are again very regular, and
the very few avoided crossings also lead to smoother behavior of the
energy levels as a function of $\alpha_R / v_{osc}$.  It is less
visible on fig.~\ref{figI}, but the energy shifts diminish with
increasing single-particle energy, as consistent with less effect on
unoccupied levels with increasing distance from the Fermi level.

In the lower panel of fig.~\ref{figI} we break the time-reversal
symmetry but stay cylindrical. The energy levels are not doubly degenerate
anymore except for $\alpha_R =0$. However, since $J_z$ is still a good quantum
number the energy levels from different total angular momentum multiplets
are allowed to cross for finite values of $\alpha_R$.
The particle number of $N=22$ only partially fill the last oscillator shell for
$\alpha_R=0$.  The picture appears to be more complicated but in fact only
due to the doubling of the visible levels. The constant energy shifts with
$\alpha_R$ and their decrease with $\varepsilon_i$ remain.

\begin{figure}
\centering
\includegraphics[width=\linewidth, height = 8.5cm]{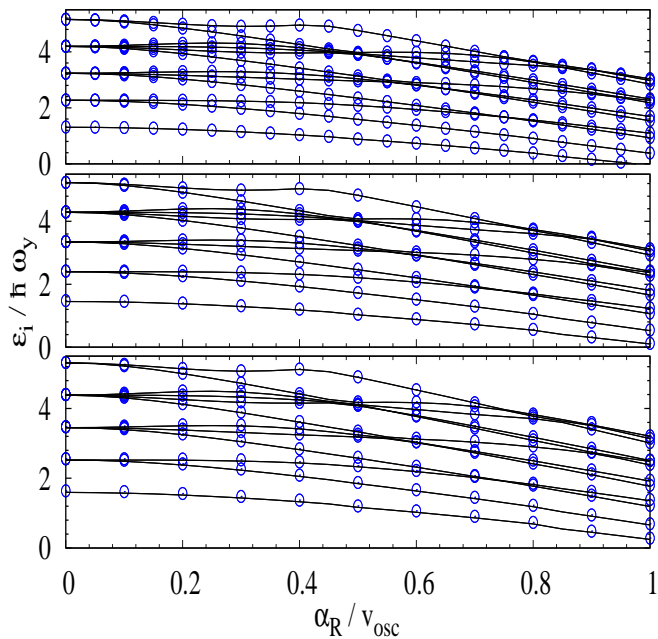}
\caption{The single-particle energy levels (divided by $\hbar \omega_y$) of interacting systems as
  function of the dimensionless spin-orbit coupling strength
  $\alpha_R / v_{osc}$.  The number of particles is $N = 20$,
  the harmonic trap is cylindrical ($\omega_x = \omega_y$), the Zeeman
  strength is $h=0$.  The blue circles are the first order perturbation
  results, and the black curves are from the Hartree-Fock solutions.
  The upper, middle and lower panels are for $g_s=0.5, 0.75, 1.0$,
  respectively. } \centering
\label{figII}
\end{figure}

To assess more directly the effect of the two-body repulsion we show
in fig.~\ref{figIa} the difference between non-interacting and
self-consistent Hartree-Fock single-particle energies as functions of
the dimensionless interaction strength $g_s$.  The dependence is
similar for the different spin-orbit strengths with the obvious
increase from zero.  The curves are denser in the lowest parts of the
figure where the largest single-particle energies appear, and thereby
demonstrating that the effect decreases with increasing energy.  In
any case, we find that all these curves increase almost precisely
linearly up to rather large strengths, $g_s$.  This strongly suggests
that perturbation also would be accurate in the same parameter range.

We can also compare the approximation~\eqref{vpert} with the difference between
the interacting and non-interacting single-particle levels. For instance,
for $N = 20$ and $g_s = 0.5$ the estimate is $V_{pert} = 0.188$ (in units of $\hbar \omega$).
From the fig.~\ref{figIa} we see that this value is rather close to the
exact value of the shift, which means that our crude approximation in the
end is quite accurate.

\begin{figure}
\centering
\includegraphics[width=\linewidth, height = 8.5cm]{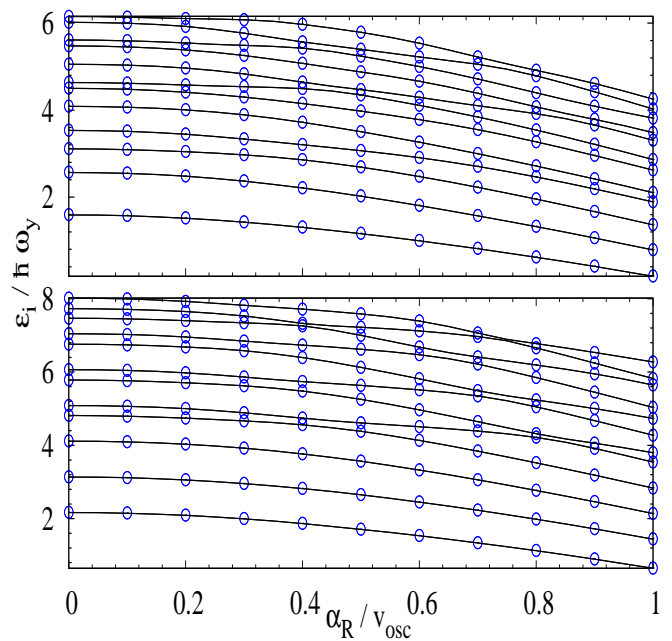}
\caption{The single-particle energy levels (divided by $\hbar \omega_y$) of interacting systems as
  function of the dimensionless spin-orbit coupling strength
  $\alpha_R / v_{osc}$.  The number of particles is $N = 20$, the interaction strength is 
  $g_s=0.5$ and the Zeeman strength is $h=0$. The upper and lower panel are for deformed 
  traps with $\omega_x = 1.57 \omega_y$ and $\omega_x = 2.71 \omega_y$. }
\centering
\label{figIII}
\end{figure}

We therefore turn to compare the Hartree-Fock spectra with the lowest
order perturbation results~\eqref{perth0}. The results are shown in
fig.~\ref{figII} for different interaction strengths, $g_s$. The
agreement is remarkably good, and the perturbation treatment is rather
accurate for all the computed values of $g_s\lesssim1$.  Deviations
increase marginally, but hardly visible on the figure, when $g_s$
approaches unity, $g_s\to 1$. Thus, the full parameter range can be
concluded to be in the weakly repulsive regime.  The perturbation
theory also provides a good approximation for deformed traps.  This
can be seen in fig.~\ref{figIII} for two non-integer frequency ratios,
and still for time-reversal symmetric systems. The interaction
strength is moderate but the agreement is very similar to the more
degenerate cases in fig.~\ref{figII}.

\begin{figure}
\centering
\includegraphics[width=\linewidth, height = 8.5cm]{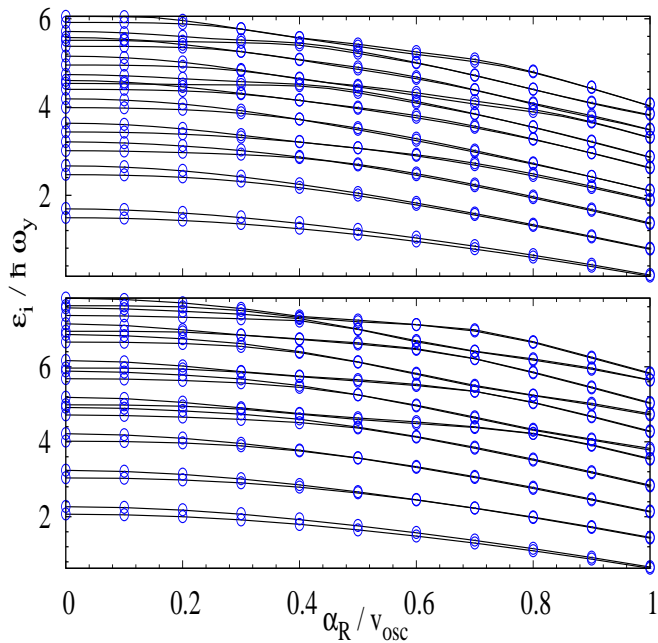}
\caption{The single-particle energy levels (divided by $\hbar \omega_y$) of interacting systems as
  function of the dimensionless spin-orbit coupling strength
  $\alpha_R / v_{osc}$.  The number of particles is $N = 20$, the interaction strength is 
  $g_s=0.5$ and the Zeeman strength is $h=0.1$. The upper and lower panels are for
  deformations $\omega_x = 1.57 \omega_y$ and
  $\omega_x = 2.71 \omega_y$, respectively. }
\centering
\label{figIV}
\end{figure}

\begin{figure}
\centering	
\includegraphics[width=\linewidth, height = 8.5cm]{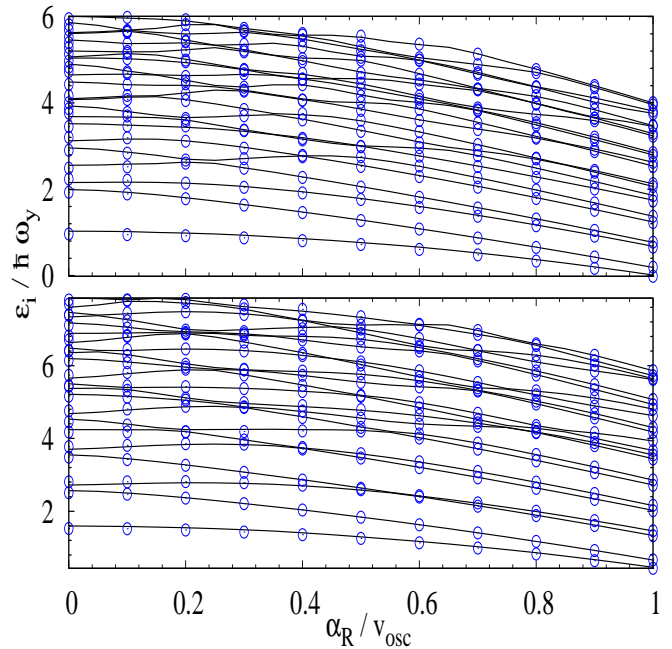}
\caption{The single-particle energy levels (divided by $\hbar \omega_y$) of interacting systems as
  function of the dimensionless spin-orbit coupling strength
  $\alpha_R / v_{osc}$.  The number of particles is $N = 20$, the interaction strength is 
  $g_s=0.5$ and the Zeeman strength is $h=0.6$. The upper and lower panels are for
  deformations $\omega_x = 1.57 \omega_y$ and
  $\omega_x = 2.71 \omega_y$, respectively. }
\centering
\label{figV}
\end{figure}

Finally, we investigate the influence of the remaining one-body
parameter, that is the Zeeman field, $h$, which lifts the
time-reversal symmetry and the corresponding degeneracy.  The
interplay between deformation, spin-orbit terms and Zeeman effect is
seen in the self-consistent solution shown in figs.~\ref{figIV} and
\ref{figV}.  Comparing to fig.~\ref{figIII}, we first notice that the
main features are maintained, except of course the lifting of the
degeneracy.  The comparison between two-body interacting and
non-interacting cases again show an overall upwards shift of all
levels in the spectra by inclusion of the repulsion.  There is also a
similar tendency of a decreasing shift with increasing
$\varepsilon_i$.

Increasing the magnetic field from $h=0.1\hbar \omega_y$,
fig.~\ref{figIV}, to $h=0.6 \hbar \omega_y$, fig.~\ref{figV},
substantially splits the previously Kramers degenerate levels. This is
most clearly seen for the lowest level which is substantially below
its previous time-reversed partner for small $\alpha_R$.  We also note
that this split decreases systematically for all levels with
increasing $\alpha_R$.  Finally, it is remarkable that the more
complicated perturbation treatment of the two-body interaction still
is fairly accurate even for relatively large $h$ and moderate to
substantial $g_s$-values.

\begin{figure}
\centering	
\includegraphics[width=\linewidth, height = 12cm]{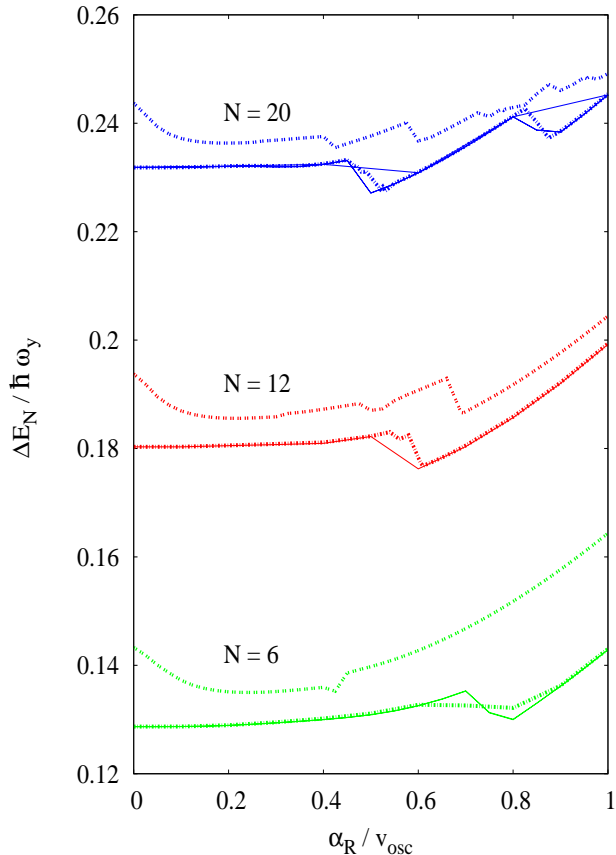}
\caption{ The total energy difference (divided by $\hbar \omega_y$) per particle as function of the
  dimensionless spin-orbit coupling strength $\alpha_R /
  v_{osc}$ from non-interacting to self-consistent Hartree-Fock
  solution.  The system is cylindrical ($\omega_x = \omega_y$) and the repulsive strength is
  $g_s=0.5$.  The number of particles are $N = 6$ (green, lower manifold), $N = 12$
  (red, middle manifold), and $N = 20$ (blue, upper manifold). The Zeeman strength is $h=0$
  (full), $h=0.1\hbar \omega_y$ (dashed), and $h=0.6\hbar \omega_y$
  (dotted).}
\label{figVI}
\end{figure}

The total energy of the system is a revealing quantity, but as we are
interested in the effects of the two-body repulsion, we prefer to show
the total energy difference between interacting and non-interacting systems per particle number
\begin{equation}
\Delta E_N = \frac{1}{N} (E_N - \sum_{j = 1}^N \varepsilon^{0}_{j}),
\end{equation} 
where $\varepsilon^{0}_{j}$ are the single-particle eigenenergies of the
non-interacting Hamiltonian and $E_N$ is the total energy of the interacting $N$-body system~\eqref{totEnergy}. 
In fig.~\ref{figVI} we present the results for a cylindrical ($\omega_x = \omega_y$) system with
moderate repulsion and varying the Zeeman field. We first notice the very
weak dependence on the spin-orbit strength for all cases.  We also see
that even when the total energy difference is divided by $N$ we see an increase
with particle number which is less than another factor of $N$.  Thus,
the energy difference is increasing with $N$ by a power between $1$ and $2$,
i.e. $\Delta E_N\propto N^\delta$, $1\leq \delta\leq 2$.
The rapid decrease of the energy differences correspond to the points of spectra
where many levels come close to each other. The small Zeeman strength 
$h=0.1\hbar\omega_y$ does not affect the total energy difference much, 
except for the points where the Fermi energy is in the vicinity of the avoided crossings. 
The level repulsion in a system with broken time-reversal symmetry is weaker and 
the additional smaller fluctuations of the total energy differences can be seen, especially for
stronger Zeeman field, e.g. $h=0.6 \hbar \omega_y$.

\begin{figure}
\centering	
\includegraphics[width=\linewidth, height = 12cm]{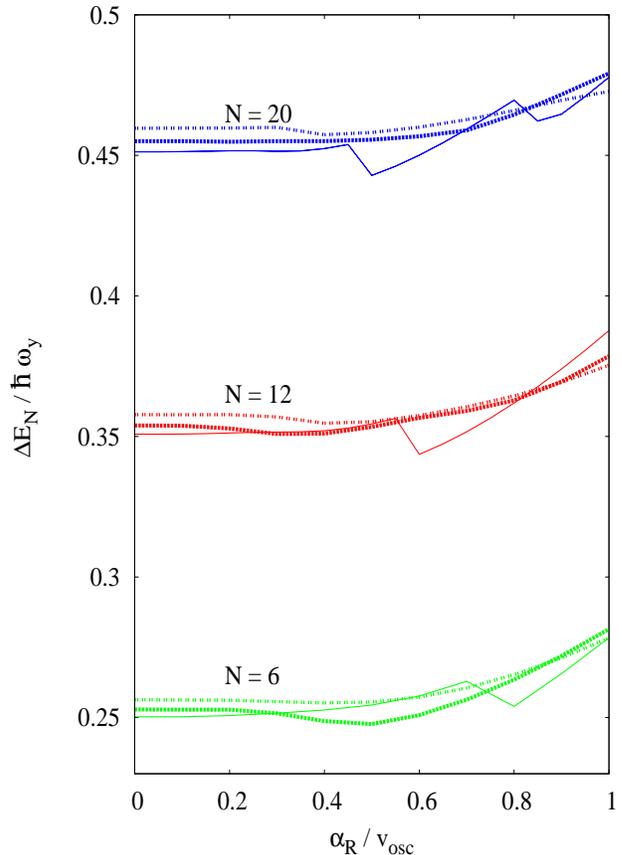}
\caption{The total energy difference (divided by $\hbar \omega_y$) per particle as function of the
  dimensionless spin-orbit coupling strength $\alpha_R /
  v_{osc}$ from non-interacting to self-consistent Hartree-Fock
  solution. The Zeeman strength is $h=0$ and the interaction strength is $g_s=1$.
 The deformations are cylindrical ($\omega_x = \omega_y$) (full), $\omega_x=
  1.57\omega_y$ (dashed), and $\omega_x= 2.71\omega_y$ (dotted).  }
\label{figVII}
\end{figure}

The energy difference is obviously increasing with the strength of the
repulsion as seen by the larger numerical values on the vertical axis
in fig.~\ref{figVII}.  The deformation dependence is in contrast very
weak for all investigated particle numbers.  In both figs.~\ref{figVI}
and ~\ref{figVII} we see a few abrupt changes of the energy at
different spin-orbit couplings depending on $N$.  They arise when two
single-particle levels avoid crossing each other at the Fermi energy.
The last occupied single-particle wave function changes structure over
a small range of coupling strengths precisely in these regions.  For a
relatively weak interaction this happens very quickly over a very
small change of spin-orbit parameter. Therefore the abrupt change
would disappear only with a very much finer grid, and this continuity
would only be invisible on a much smaller scale than exhibited on
these figures.  Other avoided crossings below the Fermi energy do not
change the total antisymmetrized product wave function in this abrupt
manner because the system populates those levels both before and after
the crossings.  The same conclusion of no abrupt changes is even more
obvious for crossings above the Fermi level since none of them can
influence the total wave function.

\subsection{Statistical analysis}

In this subsection we provide results of the statistical analysis of
the calculated spectra for the interacting system.  The spacing
distribution, $P(S)$, of the nearest neighbor energy spacing, $S$, can
provide valuable insight into the dynamical properties of the system
\cite{haake2001}.  The distribution and the nearest neighbor measure
are defined as dimensionless and scale-independent values through a rather involved
standard procedure in these statistical analyses. The distribution is
then normalized and the average level spacing is equal to unity.  The
procedure is in literature referred to as ``unfolding''.

The Poisson distribution, $P(S) = e^{-S}$, corresponds to the
classically regular behavior and the class of Wigner distributions,
$P(S) \sim S^{\beta} e^{-S^2}$, serve as a so-called quantum signature
of chaos~\cite{haake2001, berry1987}.  Here the values $\beta = 1,2,4$
corresponds to different symmetries of the system ~\cite{haake2001}.
Such a statistical analysis was performed in Ref.~\cite{marchukov2014a} for
non-interacting particles in a deformed harmonic trap where each
particle is subject to a one-body spin-orbit interaction.  It was
shown that the nearest neighbor energy spacing reproduces the
$\beta=4$ Wigner distribution,
\begin{equation}
\label{wigner4}
P(S) = \frac{2^{18} S^4}{3^6 \pi^3} \mathrm{exp} \left( - \frac{64 S^2}{9 \pi} \right),
\end{equation}
for specific values of the potential parameters. Here we investigate
how the two-body interaction affects these energy level distributions.

The statistical analysis employs a non-trivial procedure to extract
the significant scale independent distributions ~\cite{haake2001}. The
basic ingredients are the single-particle energies in an appropriate
energy interval. It is required for the analysis that the
average energy spacing is normalized to unity. This procedure is usually
called unfolding of the spectrum\cite{haake2001}. The resulting spectrum is
dimensionless and so is the nearest neighbor spacing $S$. The choice of the energy interval is
restricted to the levels that are sufficiently accurately determined.
This implies that the highest single-particle energies $\varepsilon_i$
should be avoided since they reflect properties of the basis more than
effects of the interaction.  We have therefore always chosen the upper
boundary of the energy window as less than half of the number of basis
states.

The mean-field solutions for finite two-body interaction depend on
particle number.  The dynamic behavior reflected in the nearest
neighbor energy distribution must be extracted by use of an energy
interval corresponding to the allowed excitation energies of the
entire system.  A reasonable choice is therefore to have roughly the
same number of levels above as below the Fermi energy.  This requires
that the number of particles is large enough to allow us a reasonably
accurate and well defined statistical analysis.

The self-consistent calculations for deformed potentials are time
consuming and quickly increasing with particle number.  The systematic
investigation in the previous subsections of spectra and energies as
functions of the many interaction parameters were therefore carried
out and reported for a relatively small number of particles.  For the
statistical analyses in the present subsection we have chosen to use
particle number $N=100$ for selected interaction parameters.  We used
more than 575 basis states for the diagonalization with a maximum
energy corresponding to level number $200$, that is as many above as
below the Fermi level.  The accuracies of the spectra are discussed in
section IIIA.

\begin{figure*}
\begin{tabular}{cc}
\includegraphics[scale=0.65]{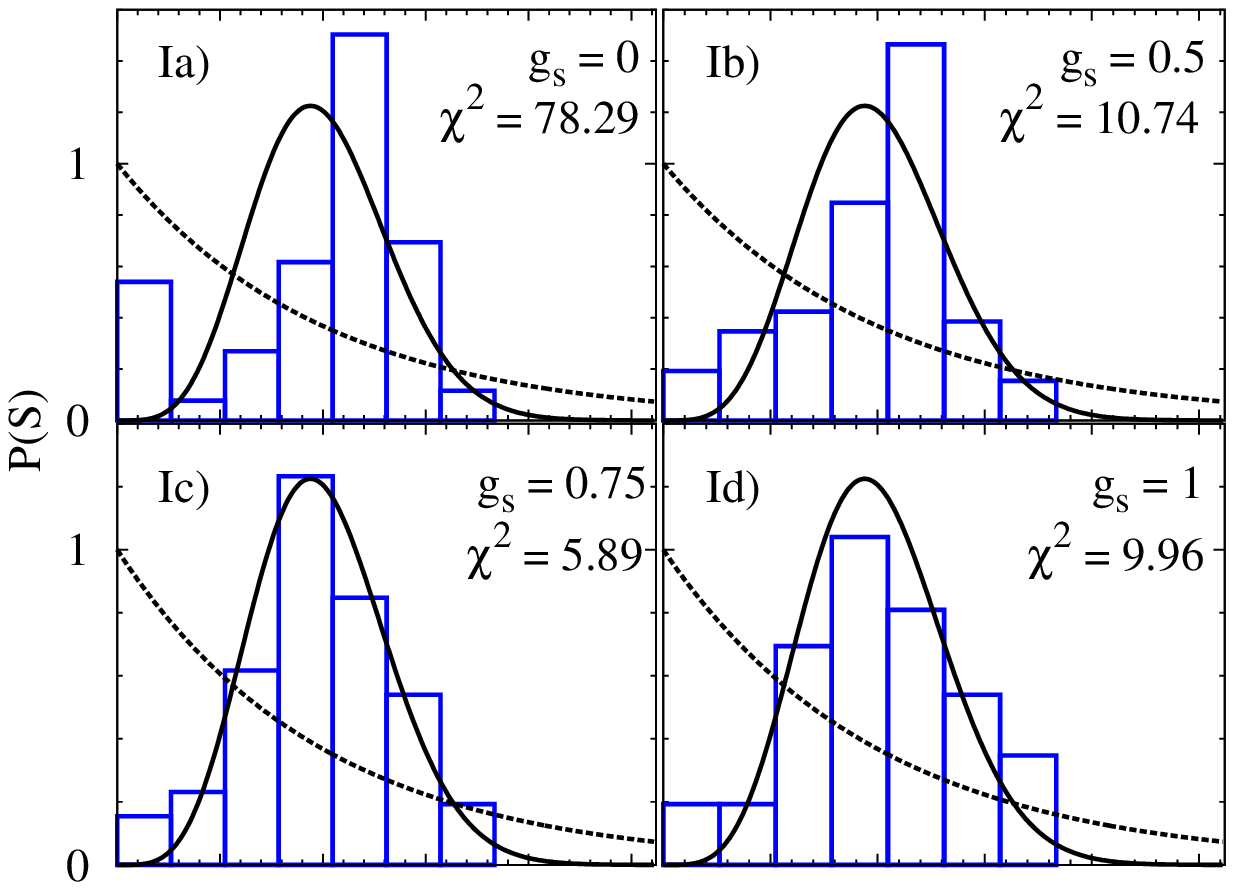} 
   & \includegraphics[scale=0.65]{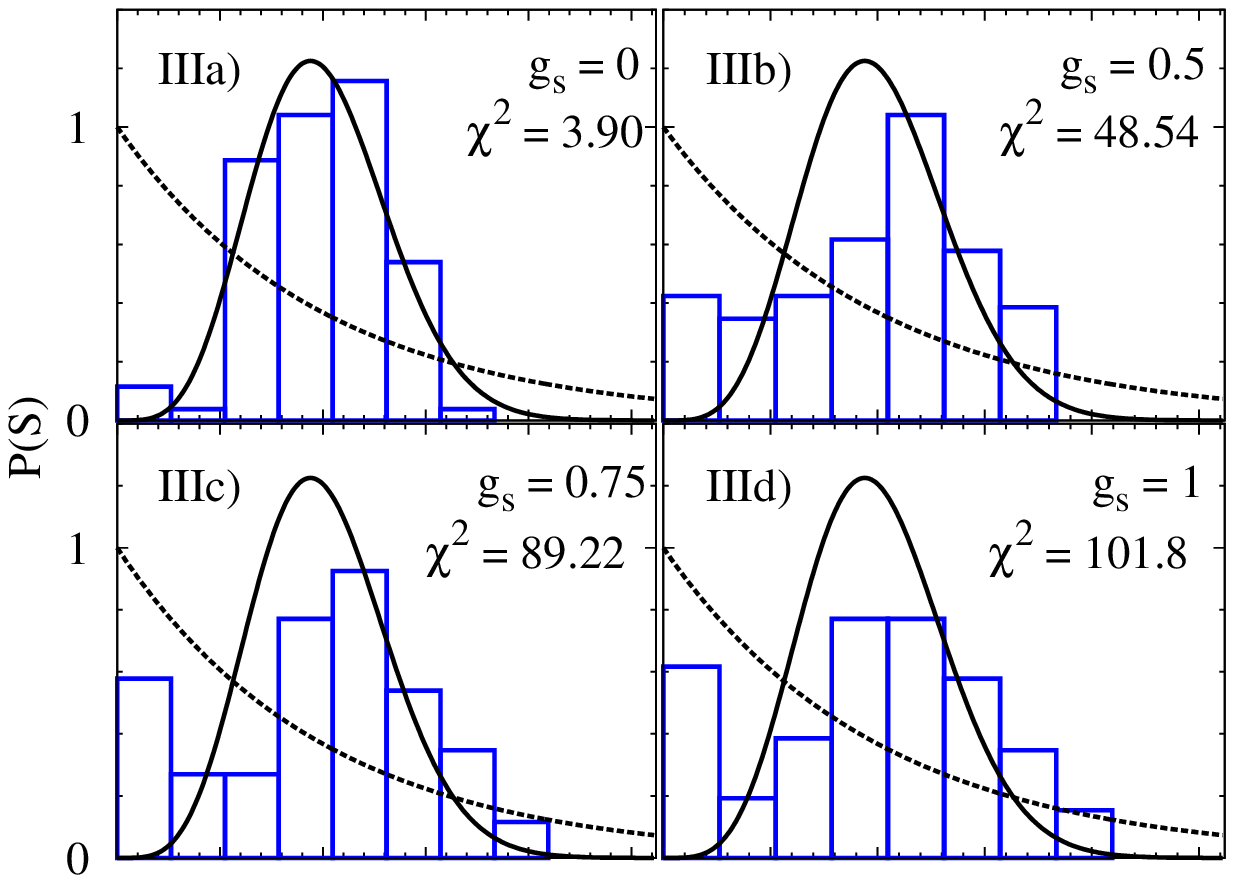}\\
\includegraphics[scale=0.65]{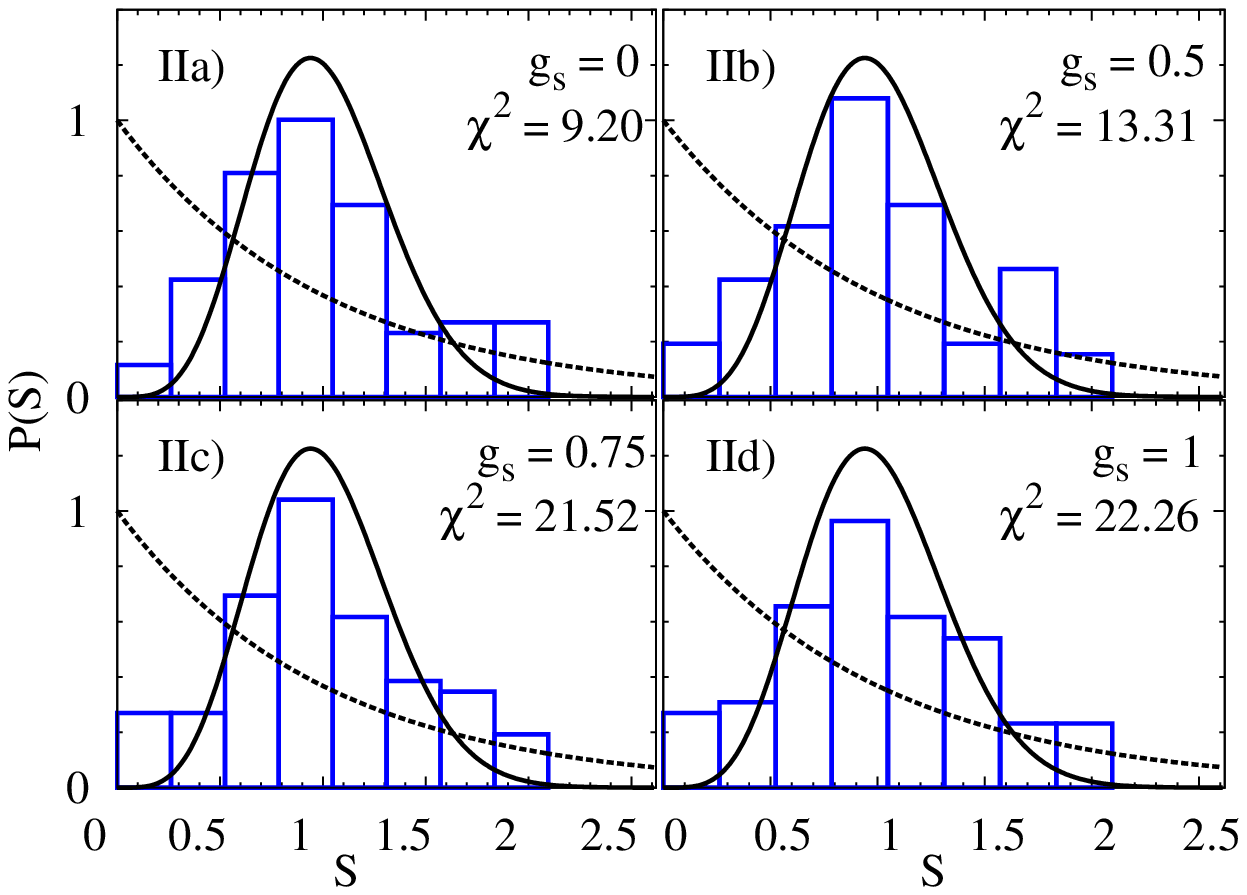} 
   & \includegraphics[scale=0.65]{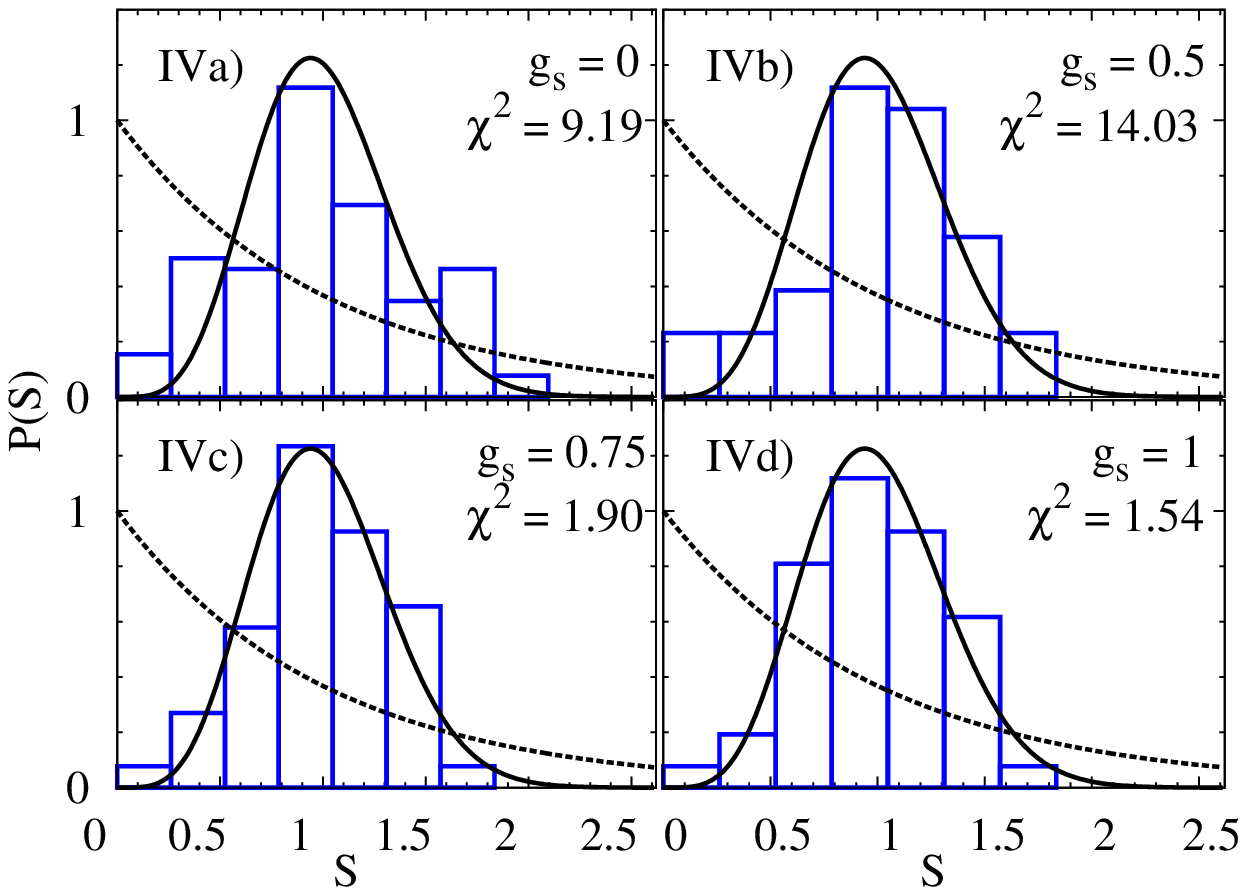}\\
\end{tabular}
\vspace*{1.0em}
\caption{The nearest neighbor level spacing distribution $P(S)$ for different sets of parameter values. Roman numerals correspond to different cases of deformation and spin-orbit coupling strength: \textrm{I}) $\omega_x/\omega_y = 1.57$, $\alpha_R/v_{osc} = 0.3$, \textrm{II}) $\omega_x/\omega_y = 1.57$, $\alpha_R/v_{osc} = 0.5$, \textrm{III}) $\omega_x/\omega_y = 2.71$, $\alpha_R/v_{osc} = 0.21$, and \textrm{IV}) $\omega_x/\omega_y = 2.71$, $\alpha_R/v_{osc} = 0.5$. The letters a-d correspond to different two-body interaction strength: a) $g_s = 0$, b) $g_s = 0.5$, c) $g_s = 0.75$, and d) $g_s = 1$. The external magnetic field is $h=0$. The distributions are compared to the Poisson (dashed line) and $\beta=4$ Wigner of eq.~\eqref{wigner4} (solid line) distributions. We used 100 non-degenerate energy levels in the analysis.}
\label{figVIII}
\end{figure*}

In fig.~\ref{figVIII} we show the distributions for two trap
deformations, $\omega_x / \omega_y = 1.57$ (\textrm{I}-\textrm{II})
with spin-orbit strengths $\alpha_R/v_{osc} = 0.3, 0.5$, respectively,
and $\omega_x / \omega_y = 2.71$ (\textrm{III}-\textrm{IV}), with
$\alpha_R/v_{osc} = 0.21, 0.5$, respectively. For each set we varied
the two-body interaction strengths $g_s$.  The deformations are chosen
to be relatively small numbers but far away from integer frequency
ratios. The values of the spin-orbit coupling strength are chosen such
that we observe distributions different from and similar to the Wigner
distribution in eq.~\eqref{wigner4}.

In fig.~\ref{figVIII}\textrm{I}) we see that for the given
deformation, $\omega_x / \omega_y = 1.57$, and spin-orbit coupling
strength, $\alpha_R = 0.3~v_{osc}$, the distribution for $g_s = 0$ is
far from eq.~\eqref{wigner4}.  However, as the repulsion increases
towards $g_s=0.75$, we notice that the Wigner distribution is
approached but for larger $g_s=1$ this similarity seems to disappear
again.  For the same deformation in fig.~\ref{figVIII}\textrm{II}),
but for larger spin-orbit strength, $\alpha_R = 0.5~v_{osc}$, the
histogram is relatively close to a Wigner distributions for $g_s=0$,
and in fact more or less conserve the character of this distribution
as $g_s$ increases.  In fig.~\ref{figVIII}\textrm{III}) the
approximate Wigner distribution at $g_s=0$ is quickly destroyed as
$g_s$ increases.  In fig.~\ref{figVIII}\textrm{IV}) we notice the
opposite effect that the Wigner distribution is much better reproduced
for $g_s=0.75$ and $g_s=1.0$ than for $g_s=0$.

These intuitive and visual qualitative evaluations of the
distributions can be made quantitative by chi-squared analyses.  The
measure of similarity between histograms and Wigner distributions can
be defined as in \cite{reichl1992}, that is
\begin{equation}
\label{chi2}
  \chi^2 = \sum_{i=1}^{i_{max}} \frac{(p_i -\nu_i)^2}{\nu_i} \;,
\end{equation}
where $i_{max}$ is a number of bins, $p_i$ is the number of the data points from the histogram in the $i$th
bin, and $\nu_i$ is the number of the data points predicted by the Wigner distribution eq.~\eqref{wigner4}.

The results are shown on each of the inserted figures in the main
fig.~\ref{figVIII}.  Fortunately, the sizes of the $\chi^2$-values
correlate perfectly with our qualitative visual observations.  When
the Wigner distribution is most precisely reproduced the
$\chi^2$-value is slightly below $2$, see \textrm{IVc} and
\textrm{IVd}.  In contrast, large deviations from the Wigner
distribution easily reach values much higher than $10$ and sometimes
significantly above $100$, see \textrm{IIIc} and \textrm{IIId}.

To reach the lowest values of about $2$ is not easily achieved with
the measure in eq.~\eqref{chi2}, since for example a small bump at
small or large $s$-values would add a rather large contribution to
$\chi^2$.  Thus, in general small deviations from the Wigner form
would show up in the $\chi^2$-value, which therefore is a sensitive
measure of an underlying chaotic dynamic behavior.

Our findings are qualitatively consistent with the results obtained in Ref.~\cite{armstrong2012} for a
many-fermion system in a cylindrical ($\omega_x = \omega_y$) 2D harmonic trap with an attractive
short-range pairing interaction. The authors of this paper have found that the interplay
between the trap and the pairing interaction is responsible for the
dynamical behavior of the system. Particularly, in the case when the contributions are of comparable
significance the competition produces the irregular behavior seen in the
Brody distribution, which is known to be an ``intermediate'' one between the
Poisson and Wigner distributions~\cite{brody1981,santos2010}. 
Similarly, in our system the dynamical behavior is defined by the interplay between the harmonic trap,
spin-orbit coupling and the two-body interaction.

\section{Conclusions}
\label{sec:conclusion}
The two-dimensional properties of spin-orbit coupled systems of
identical fermions can be investigated experimentally. It is possible
to control and independently tune several of the external one-body
parameters like deformation of an oscillator trap, the magnetic field
for Zeeman splitting, form and strength of the spin-orbit coupling,
and on top also the interaction between pairs of particles.  The
two-body interaction introduces a particle number dependence of
structure and properties.  The first attempt to understand the
interplay between all these terms of the Hamiltonian is to compute the
average structure by use of mean-field theory.

We write down the self-consistent Hartree-Fock equations with one-body
terms in the Hamiltonian.  The formulation necessarily involves a
two-component wave function formally corresponding to spin-up or
spin-down.  We discuss symmetries related to each of the one- and
two-body potentials, and relate to the properties for the Hartree-Fock
equations.  

The single-particle energy spectra exhibit the same features as the
non-interacting systems, that is with level crossings and low and high
level-density regions corresponding to more or less stability against
changes due to other interactions or other degrees of freedom.  The
overall change from the two-body repulsive potential is a shift
upwards of all spectra.  The details of these shifts are very well
accounted for by perturbation up to moderate sizes of the two-body
interaction.  

The many details of the spectra can be collected in the total energy
which is a more macroscopic quantity.  Furthermore, we eliminate most
of the overall spectral shift-dependence by comparing the energy difference due
to the repulsion as function of the various parameters.  The energy
difference increases rather strongly with the repulsive strength, that is
very crudely as the square of the strength.
The numerical results for the energy difference most often only varies
by less than about 15{\%} as function of deformation and spin-orbit
coupling strength in most cases.  In contrast, increasing the Zeeman strength strongly
suppresses the energy difference which implies that the levels actively
involved in the two-body coupling move apart, and the repulsion is
effectively reduced.  

The dynamic behavior of the interacting system is analyzed
statistically through nearest neighbor energy level distributions of the mean-field
single-particle energies. These distributions are for moderate
repulsion somewhat similar to the non-interacting system. The
Wigner distribution associated with irregular behavior turned out to 
occur for special
parameters.  When the two-body repulsion is turned on these Wigner
distributions may disappear. However, occasionally
Wigner distributions appear for intermediate strength repulsion, while
absent for both smaller and larger repulsion.

In conclusion, we have shown that the repulsive short-range two-body interaction can be
treated perturbatively for small and moderate strengths.  The spectra
are first of all shifted depending somewhat on interaction
parameters.  Statistically irregular behavior may turn into regular
behavior, and vice versa, as function of the repulsive strength.
Our results provide evidence of possible quantum chaotic 
behaviour in these spin-orbit coupled and deformed systems
also in the presence of repulsive two-body interactions of
small to intermediate strength. Cold atoms with spin-orbit
coupling could thus provide a very flexible venue for studying 
quantum chaos.

\end{document}